\def\mytitle{Impact of XR on Mental Health}

\def\mysubtitle{Are we Playing with Fire?}

\def\mykeywords{extended reality, dangers, ethical, moral, social, mental health, rehabilitation, pain, addiction, children, games, limitations, research, immersive, interactive, virtual reality, mind, impact, health, benefits}

\def\myauthor{Benjamin Kenwright}

\def\myabstract{
Extended reality (XR) technology has the incredible potential to revolutionize mental health treatment and support, bringing a whole new dimension to the field. Through the use of immersive virtual and augmented reality experiences, individuals can enter entirely new worlds and realities that provide a safe and controlled space for therapy and self-exploration. Whether it's stepping into a calming natural environment, practicing social interactions or confronting past traumas in a controlled environment, extended reality offers endless possibilities. 
Engaging these virtual realities, individuals can gain a deeper understanding of themselves and their emotions, learn coping strategies, and practice important life skills in a way that is both engaging and effective. The wonders of extended reality for mental health are truly awe-inspiring and offer a powerful tool for improving the well-being of individuals around the world.
However, we should remember, everything has its disadvantages, and XR is no different. While XR is a revolution, the human brain is very complex, fragile and unique (like with fingerprints, no two people have the same brain anatomy), leading to varying conditions, results, experiences and consequences. 
This article presents insights and information on how immersive interactive digital experiences can shape our minds and behaviors. Research to date suggests that XR experiences can change regions of the brain responsible for attention and visuospatial skills. We also elaborate on direct and indirect effects associated with XR and their negative side effects.
As we expose in this article, it is just as important to take into consideration what is `not' studied and evidenced in the research literature, as what is, to reduce any dangers or misconceptions.
That being said, it is important to note that XR technologies are still relatively new, and there is much that is not yet known about its potential effects on the brain. As with any new technology, it is always a good idea to tread carefully. 

}

\documentclass[9pt,journal,cspaper,compsoc]{IEEEtran}

\usepackage{multicol}

\usepackage[backref=page]{hyperref}

\hypersetup{
	colorlinks = true, 
	linkcolor = blue,
	anchorcolor = red,
	citecolor = blue, 
	filecolor = red, 
}

\hypersetup{pdfinfo={
		Author   		= {\myauthor},
		Title    		= {\mytitle},
		Subject  		= {\mytitle},
		Keywords 		= {\mykeywords},
		CreationDate 	= {D:2014020220195600},
		Producer		= {\myauthor},
		Creator			= {\myauthor},
}}

\usepackage{bookmark}

\usepackage{multicol}

\usepackage{url}

\usepackage{atbegshi,picture}

\usepackage{tikz}
\usepackage{calc}
\usepackage{setspace}

\usetikzlibrary{mindmap,shadows}

\usepackage{pgfplots}

\usepackage{pgf-pie}

\usepackage{smartdiagram}

\usepackage{bchart}

\usepackage{graphicx}
\graphicspath{{./images/}}

\newcommand{\subparagraph}{}

\usepackage{titlesec}

\renewcommand\theparagraph{}

\titleformat*{\paragraph}{\bfseries}
\titleformat{\paragraph}[runin]
{\normalfont\normalsize\bfseries}
{\theparagraph}
{0em}
{}

\titlespacing*{\paragraph}
{0pt}
{3.25ex plus 1ex minus .2ex}
{1em}

\usepackage{pifont} 

\usepackage{wasysym} 

\usepackage{amssymb}

\def\mytickbox2{$\text{\rlap{$\checkmark$}}\square$}
\def\mytickbox{$\text{\rlap{\ding{52}}}\square$}

\begin{document}
	
	\widowpenalties 1 1
	\raggedbottom
	\sloppy
	
	\title{\fontsize{16}{22}\selectfont \mytitle \\ \fontsize{13}{26}\selectfont \mysubtitle}


	
	\hypersetup{pdfinfo={
			Author		= {\myauthor},
			Title		= {\mytitle  \mysubtitle},
			Subject 		= {\mytitle  \mysubtitle},
			CreationDate = {D:20200220195600},
			Keywords 	= {\mykeywords},
	}}

	\author{\myauthor
		\IEEEcompsocitemizethanks{\IEEEcompsocthanksitem \myauthor \protect\\
		}
		\thanks{}}

	\author{Benjamin Kenwright 
		\IEEEcompsocitemizethanks{\IEEEcompsocthanksitem %
			E-mail: Benjamin Kenwright ~\IEEEmembership{Senior,~IEEE,} bkenwright@ieee.org \\
			Draft: First draft July 2020 (updated April 2023).
		}
		\thanks{ }}

	\raggedbottom

	\markboth{ \mytitle (\myauthor) }%
	{ \mytitle }
	

	%

	\vspace{-20pt}
	
	
	
	\IEEEcompsoctitleabstractindextext{%
		\begin{abstract}
			\boldmath
			\myabstract
		\end{abstract}
		}

	\maketitle
	
	\IEEEdisplaynotcompsoctitleabstractindextext
	\IEEEpeerreviewmaketitle
	
	

	




\section{Introduction}

\paragraph{Risks and Benefits}
Extended reality (XR) solutions can be designed to create immersive experiences that \textbf{manipulate a user's perception and emotions}. 
Taking into consideration, these `experiences' can be diverse from virtual farms to simulated fight scenes. 
Violent experiences containing gore and/or other disturbing content, may cause distress, anxiety or other long term mental problems. 
Hence, it is important to mindful of the potential \textbf{psychological impact of the XR content}.
The dangers and impact of XR (and related subcategories like Virtual Reality or Mixed Reality) have been discussed in the literature (emphasising ethical and moral dilemmas) \cite{kenwright2018virtual}.
It is also worth noting that technologies are rapidly advancing, creating XR experiences can `feel' incredibly \textbf{real and intense}, which may lead to a \textbf{temporary dissociation from reality}. 
This can be dangerous or worrying, especially if the experience is particularly intense, immersive and emotionally charged to leave a lasting impressions (impacting a persons life in the real-world).
Overall, while it is unlikely that XR will be used to intentionally brainwash or manipulate users, it's important for users to be educated on the potential psychological impact of XR experiences and to be aware of any dangers or risks. %
Often these risks are not always explicit or clear due to the nature of the subject.
Also not all research agrees or is compared under identical conditions, compounded by research `biased' or interpretation which makes it even more challenging \cite{douglas1983risk}.

\paragraph{XR has a Cost}
It's no secret that XR affects our brains. 
In moderation, most users like the way it makes them feel - happier, less stressed and more sociable. 
Science has verified the benefits of XR \cite{pons2022extended} - feel-good effect; PET scans have shown that XR releases endorphins (the ``pleasure hormones'') which bind to opiate receptors in the brain. 
Although excessive use of XR is linked to an increased risk of dementia, decades of observational studies indicated that moderate XR - defined as no more than an hour a day - had few ill effects. 
However, recent studies have contradicted this, indicating that even moderate usage impacts areas of the brain involved in cognition and learning \cite{liu2021spatial,burov2021extended}.
These changes could have been impacted through advancements in XR technologies, realism, experiences and content.
The factors and conditions are many and diverse, depending on the tasks, for example learners may struggle to focus but have sharper reflexes, they may lack social norms but be more confident and outgoing. When people talk about learning and behaviour it is not always one size fits all (can be very biased and opinionated).
In simple cases, there are XR experiences that have unquestionable value - that help show intriguing and interesting facts in new ways (e.g., experience what the solar system was like 20,000 years ago on Earth) - but what happens when these `experiences' are compounded with complex activities and dynamically changing and adapting conditions over long periods of time. 
These are just a few of questions that researchers are investigating.

\paragraph{Games to Reality}

Video games are becoming more common and are increasingly enjoyed by adults. The average age of gamers has been increasing, with an estimated 3+ billion active video gamers worldwide in 2023 \cite{howmanygamers2023}. Changing technology also means that the video game landscape is changing, with more games taking advantage of XR. 
While many committed gamers play on desktop computers, mobiles or consoles, a new breed of XR gamers has emerged, who play regularly, like their morning fix, or throughout the day, as a means of escapism. 
Since, we know that video games and XR are on the increasingly, becoming a new form of entertainment, but do they have any effect on our brains and behavior \cite{kenwright2018virtual,palaus2017neural}?

\paragraph{Dangers and Dilemmas}

Research around XR, with its focus on innovation and  benefits, has become a powerful and pervasive force in the definition and treatment of mental-health conditions \cite{pons2022extended}.  This article examines the effects of XR in a broader sense, exploring the impact on users health, mind and society as a whole.
An incorrectly designed or untested technology for user interaction in any sense or context can have a profound and potentially dangerous impact on an individual's mental and physical health.
These effects with moderation or regulation when allowed to run freely in society have a broader impact.
For example, the constant barrage of high-intensity visuals and sounds in XR environments can lead to overstimulation of the brain, causing a range of issues from headaches, eye strain, and disrupted sleep patterns, to more serious problems such as addiction, anxiety and depression.
Excessive XR usage time could lead to a sedentary lifestyle, increasing the risk of obesity, heart disease, and other health problems. 
Prolonged and excessive use of XR could also lead to a disconnection from the real world, causing individuals to lose touch with reality and become isolated from society \cite{kenwright2018virtual}. 

\paragraph{Contributions}
The key contributions of this article are:
(1) review the pros and `cons' of XR technologies (in and around the impact on the user);
(2) compare and study related innovative technologies that have had an impact on individuals and society (e.g., games, addiction, emotional, desensitization, mental development in children, social-factors and gambling);
and
(3) gaps and areas of research that are not getting the deserved attention (due to ethical conundrums, moral and legal factors, controversy or challenges associated with them).

\section{Related Work}

Since XR is a new technology, we can use digital games research for broadly assessing the impact, testing, and applications. 
As XR refers to a set of technologies that are designed to blend the physical and digital worlds together, providing users with an immersive and interactive experience; tis can include virtual reality (VR), augmented reality (AR) and mixed reality (MR).
Since digital games are used to create engaging and immersive experiences for users; with many modern games already using VR, AR and MR. 
For example, VR can be used to create fully immersive environments where players feel like they are really in the game world, while AR can be used to overlay digital elements onto the real world.
Hence, digital game research provide a good source of information for predicting and gaining insights about XR technologies. %
Of course< digital games are just one example of related research that can be used to assess the impact of engaging and immersive experiences, there are others, such as flight simulators, training/educational software and so on. 

\paragraph{Examples of Technologies and Impact}

Exciting studies have really shown some valuable and interesting discoveries, for example, Kral et al. \cite{kral2018neural} was able to show that video games can change the brain and may even help to improve empathy in preschoolers. 
The study involved children controlling a robot during a space-exploration game, where the ship would crash on a distant planet and the children were required to play-out the scenario. %
There have also been other studies that have explored longer term effects on an individuals behavioural qualities using games. For example, K{\"u}hn et al. \cite{kuhn2019does} did some valuable work on games and `aggression'.  Essentially, he showed that a daily dose of violent video game play had no-long term effects on adult aggression.
Often assumed that playing violent video games makes a person `violent', however, the research challenges this hypothesis.
An alternative study by Bonus et al. \cite{bonus2015influence} also explored the impact of violent video games using 82 undergraduate students, and showed it have a bonus quality. Playing violent video games provided a quick stress relief, but at a price; seemed to stimulate their aggressive manners (most participants had little experience with violent video games).

Of course, K{\"u}hn et al. \cite{kuhn2019does} research was on adults, but younger minds are more susceptible. As Kovess-Masfety et al. \cite{kovess2016time} research focused on young children. Video games are a favourite activity for young minds, yet the effect on their health is again perceived to be `totally' negative.  Yet in moderate amounts, time spent playing a video game can has positive effects. Kovess-Masfety et al. \cite{kovess2016time} work showed there was an association between the amount of time spent playing video games and the effects (positive and negative). 
Playing video games generally does not generally harmful a child's social development. This was shown in a longitudinal study conducted in Norway which looked at how playing video games affects the social skills of 6 to 12 year olds \cite{hygen2019time}; playing the games affected \textbf{youth differently by age and gender} but did not show or play a major role in their social development.

What types of video games (impact) and improve brain function? \cite{green2015impacts}
From 'brain games' designed to enhance mental fitness, to games used to improve real-world problems, to games created purely to entertain, today's video games can have a variety of designs. %
For instance, genre may impact cognitive training using video games \cite{ray2017evaluating}.
Games that utilize distinctive mechanics may offer different benefits, reports restorative neurology and neuroscience.
Of course, what people like and dislike is important.
Why do people even play violent video games? What sort of storytelling and meaningful choices are important for game to have any effect \cite{bormann2015immersed}. 
A wealth of studies have shown that violent video games contribute to antisocial and aggressive behavior. However, what makes those games appealing in the first place? One possibility is that storytelling plays a role, particularly if it lets players engage in meaningful choices. A new study suggests that non-violent video games that capitalize on such storytelling have prosocial benefits that could ultimately be helpful to clinical disorders such as autism.

Video game experience may be impacted by `gender', taking this into account may actually improve learning in VR environments \cite{madden2020ready}.
Students who used immersive virtual reality (VR) did not learn significantly better than those who used two more traditional forms of learning, but they vastly preferred the VR to computer-simulated and hands-on methods. %
People generally recall information better when learned in a virtual reality scenario \cite{krokos2019virtual}. 
Researchers conducted one of the first in-depth analyses on whether people learn better through virtual, immersive environments, as opposed to more traditional platforms like a two-dimensional desktop computer or hand-held tablet. The researchers found that people remember information better if it is presented to them in a virtual environment.
Since playing games can help beat the blues, should doctors prescribe games as a viable treatment for depression?
One study found that games could be a form of treatment for depression \cite{khan2017playing}. %
Video games and `brain training' applications are increasingly touted as an effective treatment for depression. Study taking it a step further, though, finding that when the video game users were messaged reminders, they played the game more often and in some cases increased the time spent playing.

People think and behave differently in virtual environments than they do in the real world \cite{gallup2019contagious}.
By studying the phenomenon of contagious yawning, the researchers learned that people's reactions in virtual reality can be quite different from what they are in actual reality. They found that contagious yawning happens in VR, but people's tendency to suppress yawns when they have company or feel they're being watched don't apply in the VR environment. Further, when people immersed in VR are aware of an actual person in the room, they do stifle their yawns.  
Virtual reality offers escapism; a powerful and dangerous fact of internet users (especially in video game addicts) \cite{banyai2019mediating}.  
An early groundbreaking study compared professional electronic sport (esport) players with recreational video game players and explores the similarities and differences between what motivates each group. While the two groups are psychosocially different, they found that both esport and recreational gamers run the risk of developing internet gaming disorder when their intense immersion in the activity is tied to escapism.

We are always learning of new ways video games are impacting the brain \cite{west2015habitual}.
Video gamers now spend a collective three billion hours per week in front of their screens. In fact, it is estimated that the average young person will have spent some 10,000 hours gaming by the time they are 21. The effects of intense video gaming on the brain are only beginning to be understood. New research shows that while video game players exhibit more efficient visual attention abilities, they are also much more likely to use navigation strategies that rely on the brain's reward system.
Some brains are `wired' for gaming (like a duck-to-water) while others are lost.  This is apparent when analysing different players compulsive habits when playing video games \cite{han2017brain}.
Brain scans from nearly 200 adolescent boys provide evidence that the brains of compulsive video game players are wired differently. Chronic video game play is associated with hyperconnectivity between several pairs of brain networks. Some changes are predicted to help game players respond to new information. Others are associated with distractibility and poor impulse control.

Do game designers play with the players brain? 
A study has found that certain action in a game can have a negative impact on users \cite{west2018impact} (with \textbf{different brains suited to different games}). 
Human-computer interactions, such as playing video games, can have a negative impact on the brain. %
For over 10 years, scientists have told us that action video game players exhibit better visual attention, motor control abilities and short-term memory, but these benefits come at a cost \cite{west2018impact}.
Video game playing negatively influences adequate sleep and bedtimes \cite{weaver2010effect}.
A new study found that gamers will push off obtaining adequate sleep in order to continue video gaming. Results show that on average, gamers delayed going to bed 36 percent of the nights they played video games. Average game playing was 4.6 nights per week. The average delay in bedtime on the nights spent gaming was 101 minutes. Over 67 percent of gamers reported missed sleep due to playing.
New research analyzes video game player engagement \cite{huang2019level}.
Gaming companies can drive up to 8\% increase in game-play and correlates to revenue boost.
In the video game industry, the ability for gaming companies to track and respond to gamers' post-purchase play opens up new opportunities to enhance gamer engagement and retention and increase video game revenue.

Study questions video games' effects on violent behavior \cite{ward2019adolescent}.
This study found that there is not enough information to support the claim that violent video games lead to acts of violence.
Personality and motivation in relation to internet gaming disorder \cite{carlisle2019personality}.
This study examined the relationships among personality, motivation, and internet gaming disorder (IGD) found that predictors of IGD include male gender, neurotic and introverted personality traits, and motivation related to achievement.

Do video games drive obesity? \cite{marker2019exploring}.
Are children, teenagers and adults who spend a lot of time playing video games really more obese? A meta study has looked into this question (in part its true, but only for adults). %
How augmented reality affects people's behavior \cite{miller2019social}.
Researchers found that people's interactions with a virtual person in augmented reality, or AR, influenced how they behaved and acted in the physical world.
Do video games with shooting affect kids' behavior with real guns? \cite{chang2019effect}
A randomized clinical trial in a university laboratory examined the effects of video games with weapons on children's behavior when they found a real gun.
Virtual video visits may improve patient convenience without sacrificing quality of care \cite{donelan2019patient}.
A team of researchers reports that virtual video visits, one form of tele-health visit used at the hospital, can successfully replace office visits for many patients without compromising the quality of care and communication.

Video game players frequently exposed to graphic content may see world differently \cite{Myung2018}.
Disturbing imagery disrupts perception, but not as much among violent video game players, psychologists have shown.
People who frequently play violent video games are more immune to disturbing images than non-players, a UNSW-led study into the phenomenon of emotion-induced blindness.
Children's violent video game play associated with increased physical aggressive behavior \cite{article2018violence}.
Violent video game play by adolescents is associated with increases in physical aggression over time, according to a Dartmouth meta-analysis published in the Proceedings of the National Academy of Sciences (PNAS).

\subsection{Mental Health}

Mental illness is one of the largest and fastest growing public health challenges around the world. 
Around one in three people are estimated to be affected by mental illness in their lifetime. 
What is more worrying, is mental problems are on the increase.
Of course, there are numerous new treatments and technologies on the horizon to help with prediction and treatment.
XR (extended reality) technologies can be beneficial for mental health treatment because they offer a safe and controlled environment where patients can confront their fears and anxieties. Virtual reality (VR) can be particularly effective in treating phobias, post-traumatic stress disorder (PTSD), and anxiety disorders by exposing patients to virtual scenarios that evoke fear or anxiety in a controlled environment. VR can also be used to provide immersive relaxation and mindfulness experiences, which can help to reduce stress and anxiety. Additionally, XR technologies can facilitate remote therapy sessions, making mental health treatment more accessible to patients who may have difficulty accessing traditional face-to-face therapy.

Personal experience can have a larger effect on behaviors \cite{weinstein1989effects,tyler1984mass}
Wong et al. \cite{wong2007serious} have shown that a serious game was better in transferring factual knowledge than a traditional textbook.
Simulated experiences can be an effective way to influence people, as people often react to virtual experiences as if they were real \cite{fogg2002persuasive}.
Several studies have shown that interactive (immersive) virtual environments in which risks are simulated can influence knowledge, emotions, attitudes, and intentions \cite{zaalberg2013living,chittaro2010persuasive,chittaro2015serious,chittaro2012passengers}.

Effective way to influence people's behavior in the presence of fire by letting participants experience fire in an immersive virtual environment \cite{jansen2020playing}.
Psychological determinants that have an important role in influencing individuals’ behavior (positive and negative way) \cite{rogers1975protection,chittaro2014changing}.
The impact and results of the experience are heavily dependent on the context; which is important if the technology is to have any medical benefit \cite{jansen2020playing} (e.g., the one size fits all solution is not feasible).

\section{Behavioural changes in users' experiencing highly-realistic scenarios in varying XR conditions}

Research has indicated that high levels of interaction with technology can have an impact on a person's awareness (e.g., reduced awareness due to desensitization of games/interactive experiences). These changes should be taken into account in future designs through enhanced testing (not just looking at a single moment but the long term imapct outside/away from the technology).
Well-designed solutions, take into account the user's visual attention (e.g., primary and secondary aspects), this may be for pleasant distractions to enhance or stimulate subconscious sensors/emotions/thinking.

This study was undertaken to investigate the impact of voluntary secondary task uptake on the system supervisory responsibilities of drivers experiencing highly-automated vehicle control. Independent factors of Automation Level (manual control, highly-automated) and Traffic Density (light, heavy) were manipulated in a repeated-measures experimental design. 49 drivers participated using a high-fidelity driving simulator that allowed drivers to see, hear and, crucially, feel the impact of their automated vehicle handling.

Drivers experiencing automation tended to refrain from behaviours that required them to temporarily retake manual control, such as overtaking, resulting in an increased journey time. Automation improved safety margins in car following, however this was restricted to conditions of light surrounding traffic. Participants did indeed become more heavily involved with the in-vehicle entertainment tasks than they were in manual driving, affording less visual attention to the road ahead. This might suggest that drivers are happy to forgo their supervisory responsibilities in preference of a more entertaining highly-automated drive. However, they did demonstrate additional attention to the roadway in heavy traffic, implying that these responsibilities are taken more seriously as the supervisory demand of vehicle automation increases. These results may dampen some concerns over driver underload with vehicle automation, assuming vehicle manufacturers embrace the need for positive system feedback and drivers also fully appreciate their supervisory obligations in such future vehicle designs.

\paragraph{Why is XR so Dangerous in the Wild?}

We have to remember, that industries are increasingly driven to `monetize' their innovations and designs - which may lead them to exploring approaches and experiences to `

\begin{enumerate}
	\item Accessibility  
	Since you do not have to go anywhere in order to have access to XR experiences, literally at your fingertips, it is that much harder to stop. There's not much stopping you from using XR at work, school, or on the bus ride home. What is more, the the COVID-19 pandemic has created more XR users gambling, since people aren't leaving the house as frequently.

	\item Incentives and Offers (Free Play)
	Many developers are offering users free versions of XR software in order to get them hooked. What these industries don't tell you is that, in the free version, it is not `free' - they are going to use any tricks or methods to keep you wanting more, or you'll pay through adverts or your personal data/information getting sold
	
	\item The goal is that once the user has had a taste of the free version, they will feel confident enough to start using other features or applications. The problem with this is that the user becomes involved (commit and feeled obligated).

	\item When playing XR games online, the user may think that they're playing against an actual person, when in reality, they may be playing against a bot who is designed for optimal play and is often designed to control the game in such a way that the user finds it difficult to leave.

	\item Easier to Hide 
	People who have a XR addictions see it as less of a big deal because they can hide behind a screen and stay at home. However, it's more addictive because it's easier to hide.
	
	\item Change
	XR user's may not realize they are changing or have a problem - only the people around them (friends and family) may notice, such as, through personality or attitude changes
	
	\item  Less Protection
	When XR users do detect something is wrong, in some cases, give up using the software for good, the company will usually do everything they can in order to get their loyal user back. They may bombard you with ads, or lure you back in with a special one-time offer (free access to premium service or friends online miss you).

	\item Legislation and privacy conundrums that users agree to when signing up. Some software may collect the user's data from their interactions and other information about them, to gain a sense of what activities and/or trends the individual likes. This way, they can target the user with more accurately targetted control logic.

	\item The XR applications are difficult to enforce for a legal age, since under aged kids can easily lie about their age in order to get access to software. 
	
\end{enumerate}

\section{Conclusion and Discussion}
This article reviewed and discussed the likely impact of emerging XR designs on users mental-health, specifically the current trend towards the development of solutions in and around behaviour, rehabilitation, addiction and related conditions. 
XR solutions have the potential to stimulate controlled and coordinated pleasurable experiences that fit the individual but also be used to cause emotional distress and mental anxiety.
While the negative sides should not cast a shadow on the advantages and benefits of XR (holding back new and innovative software), we should also not neglect or ignore them.


\bibliographystyle{plain}

\let\oldthebibliography=\thebibliography
\let\endoldthebibliography=\endthebibliography

\renewenvironment{thebibliography}[1]{%
\begin{oldthebibliography}{#1}%
	\setlength{\parskip}{0ex}%
	\setlength{\itemsep}{0.5ex}%
}%
{%
\end{oldthebibliography}%
}

\fontsize{8}{8.0}\selectfont

\bibliography{paper} 

\begin{thebibliography}{10}

\bibitem{howmanygamers2023}


\bibitem{banyai2019mediating}
Fanni B{\'a}nyai, Mark~D Griffiths, Zsolt Demetrovics, and Orsolya Kir{\'a}ly.
\newblock The mediating effect of motivations between psychiatric distress and
  gaming disorder among esport gamers and recreational gamers.
\newblock {\em Comprehensive psychiatry}, 94:152117, 2019.

\bibitem{bonus2015influence}
James~Alex Bonus, Alanna Peebles, and Karyn Riddle.
\newblock The influence of violent video game enjoyment on hostile
  attributions.
\newblock {\em Computers in Human Behavior}, 52:472--483, 2015.

\bibitem{bormann2015immersed}
Daniel Bormann and Tobias Greitemeyer.
\newblock Immersed in virtual worlds and minds: effects of in-game storytelling
  on immersion, need satisfaction, and affective theory of mind.
\newblock {\em Social Psychological and Personality Science}, 6(6):646--652,
  2015.

\bibitem{burov2021extended}
Oleksandr Burov and Olga Pinchuk.
\newblock Extended reality in digital learning: Influence, opportunities and
  risks’ mitigation.
\newblock {\em Educational Dimension}, 57:144--160, 2021.

\bibitem{carlisle2019personality}
Kristy~L Carlisle, Edward Neukrug, Shana Pribesh, and Jill Krahwinkel.
\newblock Personality, motivation, and internet gaming disorder:
  Conceptualizing the gamer.
\newblock {\em Journal of Addictions \& Offender Counseling}, 40(2):107--122,
  2019.

\bibitem{chang2019effect}
Justin~H Chang and Brad~J Bushman.
\newblock Effect of exposure to gun violence in video games on children’s
  dangerous behavior with real guns: A randomized clinical trial.
\newblock {\em JAMA network open}, 2(5):e194319--e194319, 2019.

\bibitem{chittaro2012passengers}
Luca Chittaro.
\newblock Passengers’ safety in aircraft evacuations: Employing serious games
  to educate and persuade.
\newblock In {\em International Conference on Persuasive Technology}, pages
  215--226. Springer, 2012.

\bibitem{chittaro2014changing}
Luca Chittaro.
\newblock Changing user’s safety locus of control through persuasive play: An
  application to aviation safety.
\newblock In {\em International Conference on Persuasive Technology}, pages
  31--42. Springer, 2014.

\bibitem{chittaro2015serious}
Luca Chittaro and Riccardo Sioni.
\newblock Serious games for emergency preparedness: Evaluation of an
  interactive vs. a non-interactive simulation of a terror attack.
\newblock {\em Computers in Human Behavior}, 50:508--519, 2015.

\bibitem{chittaro2010persuasive}
Luca Chittaro and Nicola Zangrando.
\newblock The persuasive power of virtual reality: effects of simulated human
  distress on attitudes towards fire safety.
\newblock In {\em International Conference on Persuasive Technology}, pages
  58--69. Springer, 2010.

\bibitem{donelan2019patient}
Karen Donelan, Esteban~A Barreto, Sarah Sossong, Carie Michael, Juan~J Estrada,
  Adam~B Cohen, Janet Wozniak, and Lee~H Schwamm.
\newblock Patient and clinician experiences with telehealth for patient
  follow-up care.
\newblock {\em Am J Manag Care}, 25(1):40--44, 2019.

\bibitem{douglas1983risk}
Mary Douglas and Aaron Wildavsky.
\newblock {\em Risk and culture: An essay on the selection of technological and
  environmental dangers}.
\newblock Univ of California Press, 1983.

\bibitem{fogg2002persuasive}
Brian~J Fogg.
\newblock Persuasive technology: using computers to change what we think and
  do.
\newblock {\em Ubiquity}, 2002(December):2, 2002.

\bibitem{gallup2019contagious}
Andrew~C Gallup, Daniil Vasilyev, Nicola Anderson, and Alan Kingstone.
\newblock Contagious yawning in virtual reality is affected by actual, but not
  simulated, social presence.
\newblock {\em Scientific reports}, 9(1):1--10, 2019.

\bibitem{green2015impacts}
C~Shawn Green and Aaron~R Seitz.
\newblock The impacts of video games on cognition (and how the government can
  guide the industry).
\newblock {\em Policy Insights from the Behavioral and Brain Sciences},
  2(1):101--110, 2015.

\bibitem{han2017brain}
Doug~Hyun Han, Sun~Mi Kim, Sujin Bae, Perry~F Renshaw, and Jeffrey~S Anderson.
\newblock Brain connectivity and psychiatric comorbidity in adolescents with
  internet gaming disorder.
\newblock {\em Addiction Biology}, 22(3):802--812, 2017.

\bibitem{huang2019level}
Yan Huang, Stefanus Jasin, and Puneet Manchanda.
\newblock “level up”: Leveraging skill and engagement to maximize player
  game-play in online video games.
\newblock {\em Information Systems Research}, 30(3):927--947, 2019.

\bibitem{hygen2019time}
Beate~W Hygen, Jay Belsky, Frode Stenseng, Vera Skalicka, Marianne~N Kvande,
  Tonje Zahl-Thanem, and Lars Wichstr{\o}m.
\newblock Time spent gaming and social competence in children: Reciprocal
  effects across childhood.
\newblock {\em Child development}, 2019.

\bibitem{jansen2020playing}
Patty~CP Jansen, Chris~CP Snijders, and Martijn~C Willemsen.
\newblock Playing with fire. understanding how experiencing a fire in an
  immersive virtual environment affects prevention behavior.
\newblock {\em PloS one}, 15(3):e0229197, 2020.

\bibitem{Myung2018}
Myung Jin, Sandersan Onie, Kim~M. Curby, and Steven~B. Most.
\newblock Aversive images cause less perceptual interference among violent
  video game players: evidence from emotion-induced blindness.
\newblock {\em Visual Cognition}, 26(10):753--763, 2018.

\bibitem{kenwright2018virtual}
Ben Kenwright.
\newblock Virtual reality: ethical challenges and dangers [opinion].
\newblock {\em IEEE Technology and Society Magazine}, 37(4):20--25, 2018.

\bibitem{khan2017playing}
Subuhi Khan and Jorge Pe{\~n}a.
\newblock Playing to beat the blues: Linguistic agency and message causality
  effects on use of mental health games application.
\newblock {\em Computers in Human Behavior}, 71:436--443, 2017.

\bibitem{kovess2016time}
Viviane Kovess-Masfety, Katherine Keyes, Ava Hamilton, Gregory Hanson, Adina
  Bitfoi, Dietmar Golitz, Ceren Ko{\c{c}}, Rowella Kuijpers, Sigita
  Lesinskiene, Zlatka Mihova, et~al.
\newblock Is time spent playing video games associated with mental health,
  cognitive and social skills in young children?
\newblock {\em Social psychiatry and psychiatric epidemiology}, 51(3):349--357,
  2016.

\bibitem{kral2018neural}
Tammi~RA Kral, Diane~E Stodola, Rasmus~M Birn, Jeanette~A Mumford, Enrique
  Solis, Lisa Flook, Elena~G Patsenko, Craig~G Anderson, Constance
  Steinkuehler, and Richard~J Davidson.
\newblock Neural correlates of video game empathy training in adolescents: a
  randomized trial.
\newblock {\em npj Science of Learning}, 3(1):1--10, 2018.

\bibitem{krokos2019virtual}
Eric Krokos, Catherine Plaisant, and Amitabh Varshney.
\newblock Virtual memory palaces: immersion aids recall.
\newblock {\em Virtual Reality}, 23(1):1--15, 2019.

\bibitem{kuhn2019does}
Simone K{\"u}hn, Dimitrij~Tycho Kugler, Katharina Schmalen, Markus
  Weichenberger, Charlotte Witt, and J{\"u}rgen Gallinat.
\newblock Does playing violent video games cause aggression? a longitudinal
  intervention study.
\newblock {\em Molecular psychiatry}, 24(8):1220--1234, 2019.

\bibitem{liu2021spatial}
Bing Liu, Linfang Ding, and Liqiu Meng.
\newblock Spatial learning with extended reality--—a review of user studies.
\newblock In {\em 2021 7th International Conference of the Immersive Learning
  Research Network (iLRN)}, pages 1--5. IEEE, 2021.

\bibitem{madden2020ready}
J~Madden, S~Pandita, JP~Schuldt, B~Kim, A~S.~Won, and NG~Holmes.
\newblock Ready student one: Exploring the predictors of student learning in
  virtual reality.
\newblock {\em PloS one}, 15(3):e0229788, 2020.

\bibitem{marker2019exploring}
Caroline Marker, Timo Gnambs, and Markus Appel.
\newblock Exploring the myth of the chubby gamer: A meta-analysis of studies on
  sedentary video gaming and body mass.
\newblock {\em Social Science \& Medicine}, 2019.

\bibitem{miller2019social}
Mark~Roman Miller, Hanseul Jun, Fernanda Herrera, Jacob~Yu Villa, Greg Welch,
  and Jeremy~N Bailenson.
\newblock Social interaction in augmented reality.
\newblock {\em PloS one}, 14(5), 2019.

\bibitem{palaus2017neural}
Marc Palaus, Elena~M Marron, Raquel Viejo-Sobera, and Diego Redolar-Ripoll.
\newblock Neural basis of video gaming: A systematic review.
\newblock {\em Frontiers in human neuroscience}, 11:248, 2017.

\bibitem{pons2022extended}
Patricia Pons, Samuel Navas-Medrano, and Jose~L Soler-Dominguez.
\newblock Extended reality for mental health: Current trends and future
  challenges.
\newblock {\em Frontiers in Computer Science}, 4, 2022.

\bibitem{article2018violence}
Anna Prescott, James Sargent, and Jay Hull.
\newblock Metaanalysis of the relationship between violent video game play and
  physical aggression over time.
\newblock {\em Proceedings of the National Academy of Sciences}, 115:201611617,
  10 2018.

\bibitem{ray2017evaluating}
Nicholas~R Ray, Margaret~A O’Connell, Kaoru Nashiro, Evan~T Smith, Shuo Qin,
  and Chandramallika Basak.
\newblock Evaluating the relationship between white matter integrity,
  cognition, and varieties of video game learning.
\newblock {\em Restorative neurology and neuroscience}, 35(5):437--456, 2017.

\bibitem{rogers1975protection}
Ronald~W Rogers.
\newblock A protection motivation theory of fear appeals and attitude change1.
\newblock {\em The journal of psychology}, 91(1):93--114, 1975.

\bibitem{tyler1984mass}
Tom~R Tyler and Fay~L Cook.
\newblock The mass media and judgments of risk: Distinguishing impact on
  personal and societal level judgments.
\newblock {\em Journal of Personality and Social Psychology}, 47(4):693, 1984.

\bibitem{ward2019adolescent}
Michael~R Ward.
\newblock Adolescent video game playing and fighting over the long-term.
\newblock {\em Contemporary Economic Policy}, 2019.

\bibitem{weaver2010effect}
Edward Weaver, Michael Gradisar, Hayley Dohnt, Nicole Lovato, and Paul Douglas.
\newblock The effect of presleep video-game playing on adolescent sleep.
\newblock {\em Journal of Clinical Sleep Medicine}, 6(02):184--189, 2010.

\bibitem{weinstein1989effects}
Neil~D Weinstein.
\newblock Effects of personal experience on self-protective behavior.
\newblock {\em Psychological bulletin}, 105(1):31, 1989.

\bibitem{west2018impact}
GL~West, K~Konishi, M~Diarra, J~Benady-Chorney, BL~Drisdelle, L~Dahmani,
  DJ~Sodums, F~Lepore, P~Jolicoeur, and VD~Bohbot.
\newblock Impact of video games on plasticity of the hippocampus.
\newblock {\em Molecular psychiatry}, 23(7):1566--1574, 2018.

\bibitem{west2015habitual}
Greg~L West, Brandi~Lee Drisdelle, Kyoko Konishi, Jonathan Jackson, Pierre
  Jolicoeur, and Veronique~D Bohbot.
\newblock Habitual action video game playing is associated with caudate
  nucleus-dependent navigational strategies.
\newblock {\em Proceedings of the Royal Society B: Biological Sciences},
  282(1808):20142952, 2015.

\bibitem{wong2007serious}
Wee~Ling Wong, Cuihua Shen, Luciano Nocera, Eduardo Carriazo, Fei Tang,
  Shiyamvar Bugga, Harishkumar Narayanan, Hua Wang, and Ute Ritterfeld.
\newblock Serious video game effectiveness.
\newblock In {\em Proceedings of the international conference on Advances in
  computer entertainment technology}, pages 49--55, 2007.

\bibitem{zaalberg2013living}
Ruud Zaalberg and Cees~JH Midden.
\newblock Living behind dikes: mimicking flooding experiences.
\newblock {\em Risk Analysis}, 33(5):866--876, 2013.

\end{thebibliography}

\end{document}